\documentclass[12pt]{article}

\usepackage{graphicx}
\usepackage{times}

\newenvironment{sciabstract}{%
\begin{quote} \bf}
{\end{quote}}

\newcounter{lastnote}

\title{A quantum cascade phonon-polariton laser}

\author
{Keita Ohtani, Camille Ndebeka-Bandou, Lorenzo Bosco, Mattias Beck, J\'er\^ome Faist\\
\\
\normalsize{ETH Zurich, Institute of Quantum Electronics, Auguste-Piccard-Hof 1, Zurich 8093, Switzerland}
\\
\normalsize{E-mail: jerome.faist@phys.ethz.ch}
}

\date{}

\begin{document}

\baselineskip24pt

\maketitle

\begin{sciabstract}
We report a laser that coherently emits phonon-polaritons, quasi-particles arising from the coupling between photons and transverse optical phonons. The gain is provided by an intersubband transition in a quantum cascade structure. The polaritons at h$\nu$ = 45.4meV (corresponding to an emission frequency of 10.4THz) are formed by the transverse AlAs phonon mode of monolayer thin AlInAs layers coupled to the optical modes of a Fabry-Perot cavity. The frequency location of the laser mode is in good agreement with the computed polaritonic dispersion and allows to quantify the constituent fractions of the emitted polaritons that reach a maximum of 50\% for the phonon fraction. A fraction of the gain (between 2-5\%) originates directly from the coupling between the intersubband and the phonon polarizations. The device exhibits a very low temperature dependence of its threshold current, as well as the capability to operate in very thin ($\lambda/20$) optical cavities.
\end{sciabstract}

The phonons, the quantized vibrational modes of a crystal, are by nature bosonic. It is therefore legitimate to consider a laser-like process in which a coherent population of a mechanical excitation of a solid with an occupation number much larger than one is created and maintained by pumping. The first implementation of such an idea used as phonon modes the mechanical excitations of the Mg$^+$ ion in a trap potential~\cite{VAHALA:2009ih}. Using an opto-mechanical  platform, a phonon laser operating at a phonon frequency of 21MHz was demonstrated by optical pumping using a Raman-like process in a pair of whispering gallery microresonators~\cite{Grudinin:2010p1495}. A pure mechanical phonon laser was achieved recently~\cite{Mahboob:2013dz} using a piezo-electric excitation of a micromechanical resonator.  This device, operating at frequencies of 100kHz was described as being the acoustic analog to a Brillouin laser. Amplification of acoustic phonons at 440GHz in a biased GaAs/AlGaAs superlattice was reported recently~\cite{Beardsley:2010de}.

However, it would be especially interesting to consider such lasers operating at much higher frequencies, corresponding to the Longitudinal Optical (LO) phonons frequency of solid state materials. Such oscillators, operating in the THz frequencies would be welcomed in this frequency range that is plagued by a lack of convenient optical sources~\cite{Tonouchi:2007p1411}. Initial theoretical considerations envisaged, besides phonons, the plasma waves excitations of polar semiconductors~\cite{Wolff:1970uk}. Experimentally, the observation of spontaneous emission was reported in a narrow-gap HgCdTe alloy in the regime where the gap was resonant with the LO phonon energy~\cite{Fuchs:1991tw}. However, as was pointed out  by Chen and Khurgin~\cite{JingChen:2003jh}, the very large density of states of the LO phonons makes the realization of such a LO phonon laser difficult. Indeed, measurements of the optical phonon occupation by Raman scattering in Quantum Cascade Lasers structures showed that the distribution remained very close to a thermal one especially when the temperature is raised~\cite{Spagnolo:2002p1956}.

\begin{figure}[h]
  \centering
  \includegraphics[width=9cm]{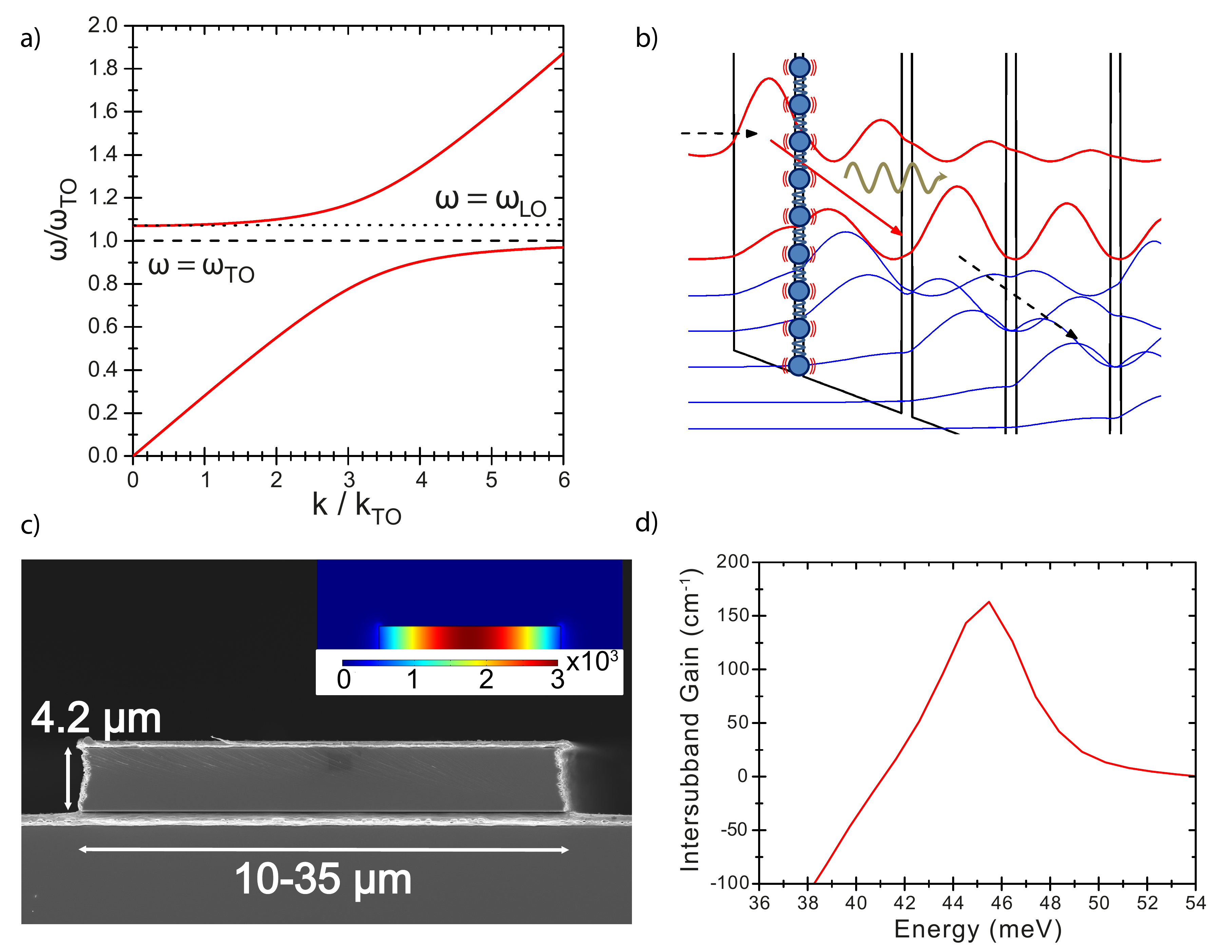}
  \caption{ a) Schematic dispersion of a phonon-polariton, showing the two characteristic frequencies of the LO and TO phonons as well as the region where the mixed particle arise. b) Schematic plot of the heterostructure potential and square modulus of the wavefunction for the phonon-polariton laser EP1531. As shown schematically, the phonon mode is contained within the monolayer thick AlInAs. The gain arises from the population inversion occurring between the upper and lower states (in red) maintained by electrical injection (indicated dashed arrow). c) Polaritonic waveguide, using two metal layers enclosing the active region in order to confine the optical part of the polariton. The computed square modulus of the electric field is plotted in the inset using a color scale.  d) Computation of the intersubband gain of the device. }
  \label{fig:bandstructure}
\end{figure}

In this work, we propose that coupling optical phonons excitations to light, effectively creating phonon-polaritons, alleviates this problem and enables the operation of a phonon-polariton laser. Indeed, as shown in Fig.~\ref{fig:bandstructure} a) where the dispersion of such polaritons is reproduced, the ionic polarization of the polar phonon couples to the light, creating two branches separated by a polaritonic gap~\cite{Henry:1965tz}. The two asymptotes of the dispersion (for $k \rightarrow 0$ and $k \rightarrow \infty$) correspond to the LO and TO phonon energies. In the region where the two dispersions anticross, the coupling of the phonon modes to the light effectively decreases the density of states  effectively lowering the threshold current density for a polariton laser. 

A microcavity polariton is formed through the hybridization of an interband exciton with the optical mode of a microcavity. Such an excitation was first observed experimentally by Weisbuch and Arakawa~\cite{Weisbuch:1992p12} in a III-V microcavity. Polariton lasers were described as having properties that were intermediate between a photon laser and a Bose-Einstein condensation of excitons~\cite{Imamoglu:1996un}. A polariton laser, was then demonstrated using optical pumping in GaAs-based~\cite{Bajoni:2008ef} or GaN-based~\cite{Christmann:2008p1820} structures. An electrically pumped polariton laser has been recently reported under cryogenic cooling and under application of a 5T magnetic field~\cite{Schneider:2013ix}. The light mass of polaritons enabled also their Bose-Einstein condensation~\cite{Kasprzak:2006p721}. In such a condensate, the interaction between polaritons is responsible for the rich physical properties of this system, justifying a large body of experimental and theoretical work devoted to their study~\cite{Carusotto:2013gh}.

%

As shown schematically in Fig.~\ref{fig:bandstructure}b)  the phonon-polaritons were created in our structure by periodically inserting a nominal monolayer-thick AlInAs barrier in a quantum cascade laser~\cite{Faist:1994p1420} based on a InGaAs/GaAsSb heterostructure lattice matched on InP~\cite{Deutsch:2010p1906}.
The heterostructure, following a bound-to-continuum design is designed to provide a population inversion and a large  gain to the photon fraction of the polariton at a frequency corresponding to the AlAs LO phonon energy. This is confirmed by the results of the gain computation performed at a field of 19kV/cm, corresponding to a current density of 4.1kA/cm$^2$ by our density-matrix based simulator~\cite{Terazzi:2010p1513} and shown in Fig.~\ref{fig:bandstructure}d). As the AlInAs layers are at a location where the upper and lower states wavefunctions have a large amplitude, an additional gain is also expected to arise by the interaction between the intersubband and phonon polarizations. 
Compared to the exciton-polariton laser, our phonon-polariton laser is intrinsically different because the polaritons are stimulated by their coupling through both their ``photonic" and ``phononic" parts whereas the coherent state of exciton polaritons is generated by final state stimulation of exciton-phonon interaction.

The 4.3$\mu$m thick structure EP1531 was grown by molecular beam epitaxy (MBE) and comprised 60 repetitions of the active region preceded and finished by contact layers. Along with the phonon-polariton laser, a reference structure  EP1520 was also grown which was in all point identical except for the replacement of the AlInAs barrier by a GaAsSb one. As shown in Fig.~\ref{fig:bandstructure}c), waveguiding of the polariton is done in a hybrid way, as the phonon excitation is naturally confined within the AlInAs layer, while the photonic part is guided along the plane of the layers by a double metal waveguide~\cite{Unterrainer:2002ii}.  The devices were then processed by metalization, wafer bonding and substrate removal before the fabrication of the  ridges by dry etching. After soldering and wire bonding, the chips were measured in a liquid Helium flow-cryostat. 

\begin{figure}[h]
  \centering
  \includegraphics[width=7cm]{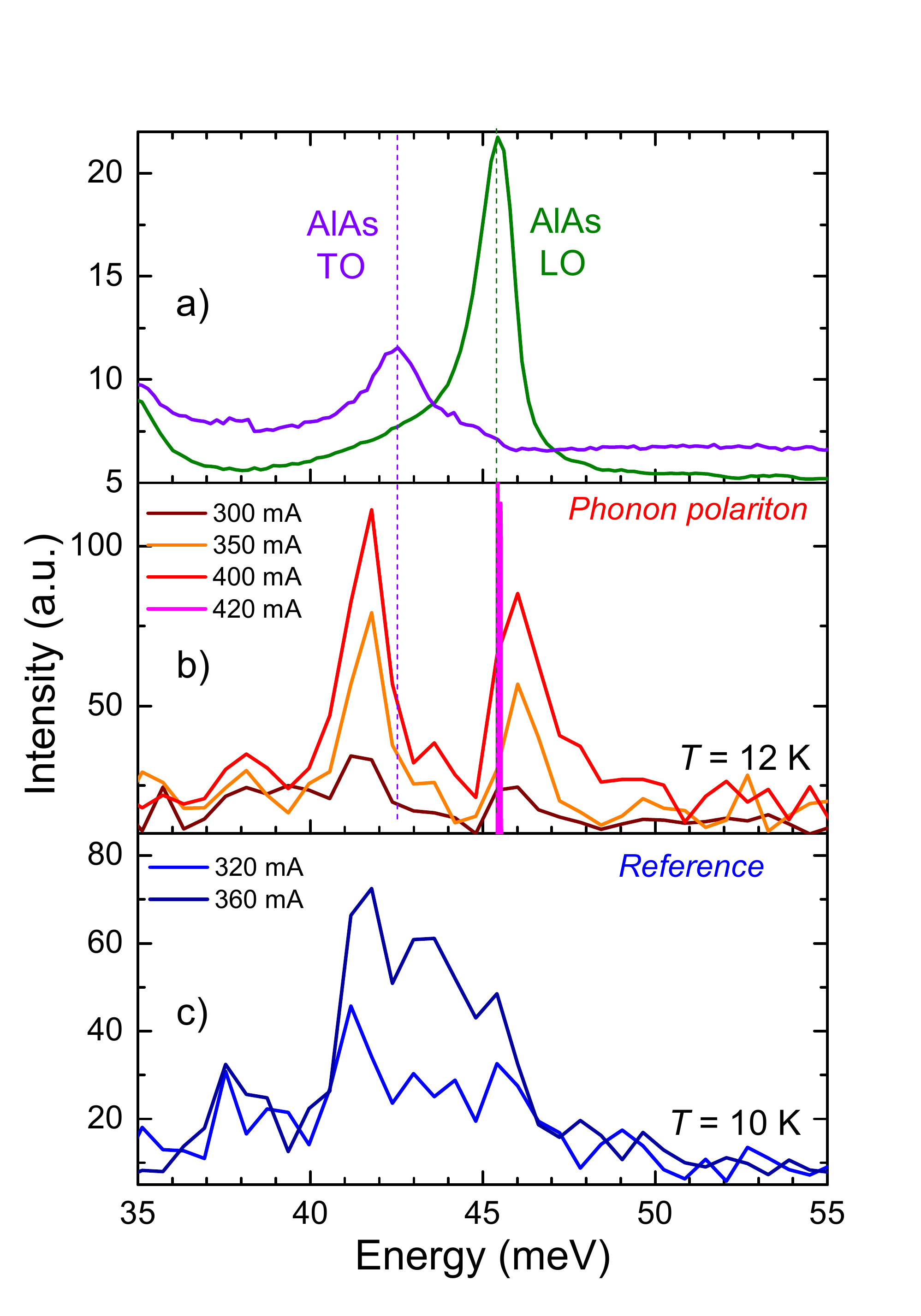}
  \caption{ a) Measurements of the LO and TO phonon frequencies in a bulk AlInAs epitaxial layer by Raman spectroscopy in the backscattering configuration. b) Subthreshold emission electroluminescence spectrum of the phonon-polariton laser structures for increasing currents, as indicated. At a current of 420mA, the laser is above threshold and the linewidth is limited by the resolution of the spectrometer. c) Same subthreshold measurements, but performed using the reference structure.}
  \label{fig:electro}
\end{figure}

The subthreshold emission spectrum of the phonon-polariton laser, recorded using a home-built vacuum FTIR fitted with a He-cooled bolometer, is compared to the one of the reference structure in Fig.~\ref{fig:electro}. While the emission spectrum of the reference structure displays a spectrum centered at 43meV with a full width at half maximum of 4meV, close to the theoretical prediction (3.5meV) shown in Fig.~\ref{fig:bandstructure}d), the spectrum of the phonon-polariton laser is asymmetric, with a broad minimum corresponding to the restrahlen band of AlAs and a sharp maximum close to the AlAs LO phonon energy. 
 %

\begin{figure}[h]
  \centering
  \includegraphics[width=10cm]{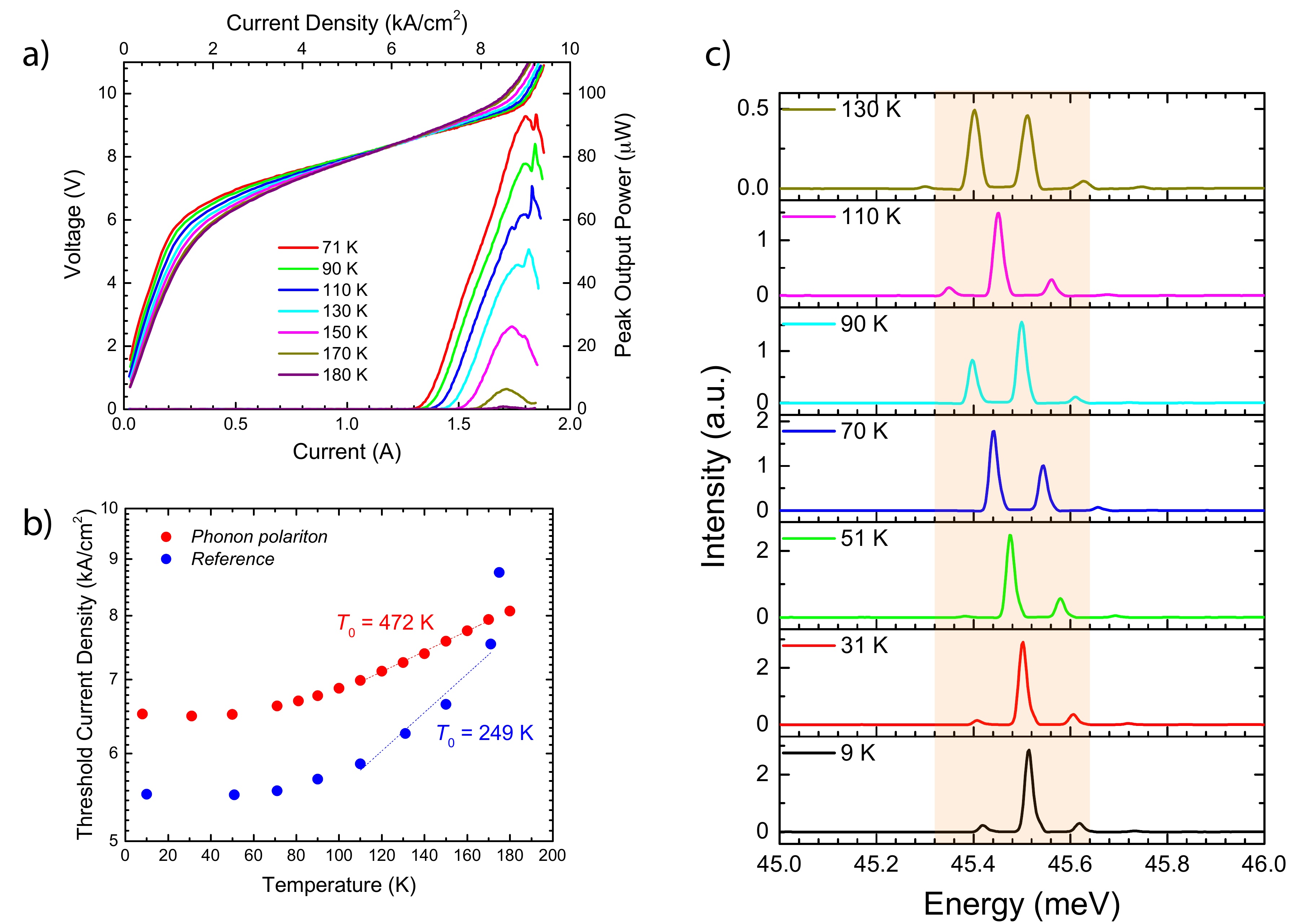}
  \caption{a) Light-current and voltage-current characteristics of the phonon-polariton laser for various temperatures between 10 and 180K, as indicated. b) Threshold current density as a function of temperature. The phonon-polariton laser exhibits a slightly higher maximum operating temperature (182 instead of 175K) and a much weaker dependence of the threshold current as a function of temperature compared to the reference structure. c) High resolution spectra of the phonon-polariton laser as a function of temperature. The main emission remains in a small energy range about 0.2meV wide (shown by the hatched region).}
  \label{fig:licomp}
\end{figure}

The light and voltage versus current characteristics of the device were measured as a function of temperature in pulsed operation and displayed in Fig.~\ref{fig:licomp} a). The peak optical power is 100$\mu$W and the device operates up to 182K. As shown in Fig.~\ref{fig:licomp} b), the temperature dependence of the threshold current can be fitted by the empirical exponential form $J_{th} = J_0 \exp ( T/T_0)$.  The characteristics temperature $T_0 = 472K$ observed for the phonon laser structure is very high, expressing an unusually small temperature dependence of the threshold current as compared to conventional quantum cascade laser structures.

A better insight in the characteristics of this laser can be gained by studying the spectral characteristics of the laser. Spectra of the structures were both measured using a high-resolution Fourier transform infrared spectrometer Brucker 80V and a DTGS detector. To minimize the broadening of the spectra, the electrical pulse length was kept below 50ns.  As shown in Fig.~\ref{fig:licomp} c), unlike in conventional quantum cascade lasers, the modes do not significantly shift with temperature in the whole operation temperature range of the device.

\begin{figure}[h]
  \centering
  \includegraphics[width=8cm]{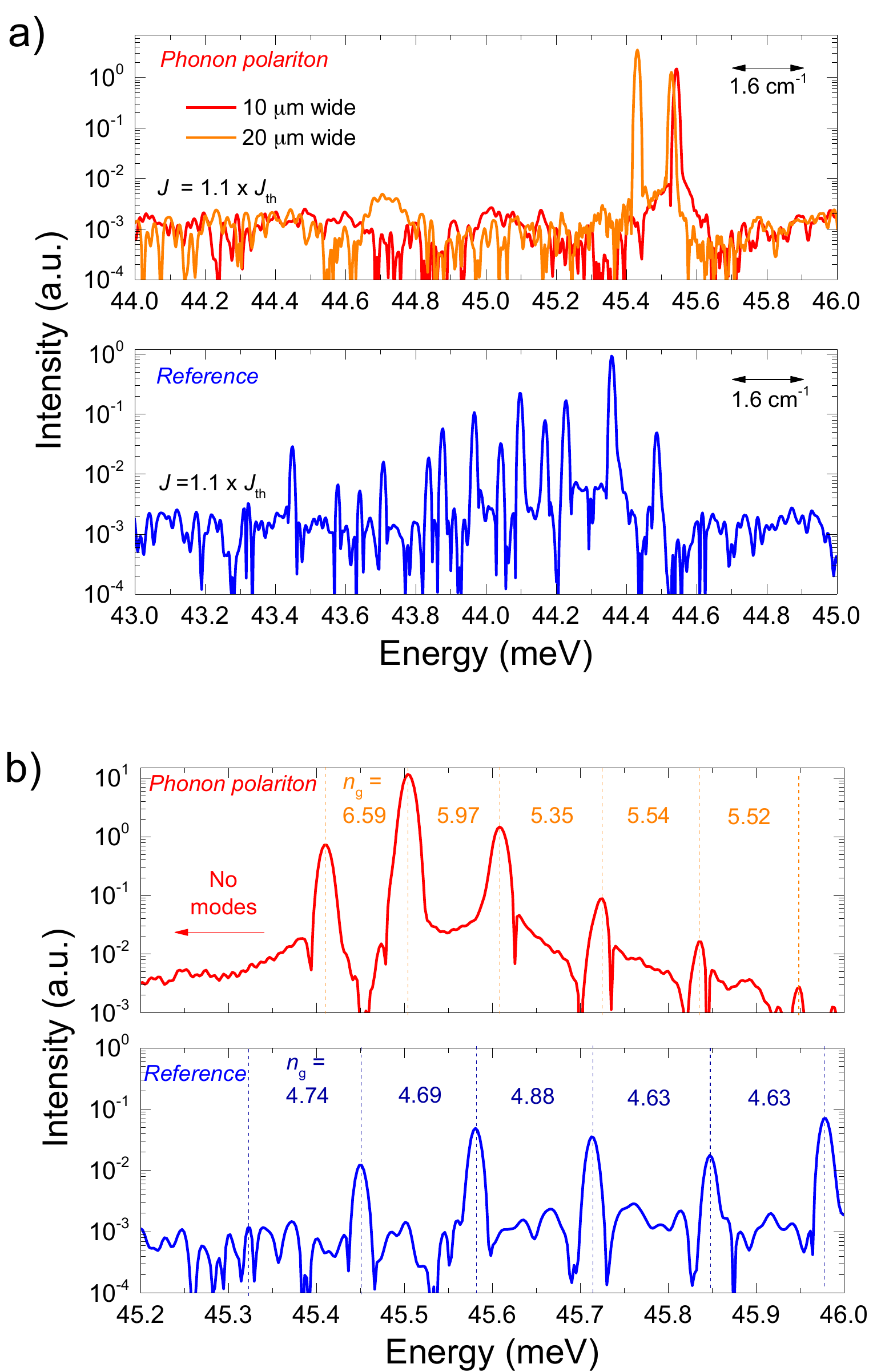}
  \caption{a) Comparison of the emission spectra of the phonon-polariton and of the reference structure at current 10\% above threshold, showing the much narrower spectrum of the phonon-polariton structure. b) Same comparison, but performed when the device is driven at its maximum power. In contrast to the reference structure, the change in spacing of the mode of the polariton laser is a testimony of the very strong dispersion of the phonon-polariton modes.}
  \label{fig:spectra}
\end{figure}

Furthermore, as shown in Fig.~\ref{fig:spectra} a)  the spectra of both phonon-polariton and reference devices, measured and compared at an injection current density $J=1.1\times J_{th}$ times the threshold value $J_{th}$ are qualitatively very different. While the spectrum of the reference structure shows laser modes over a relatively broad spectral range, the emission spectrum of the phonon-polariton laser is much narrower and typically contains one to two modes. 
%

Another important feature of the emission of the phonon-polariton laser is the characteristic dispersion of the emission. In fact, the group index $n_g$, as retrieved experimentally from the angular frequency spacing of the longitudinal modes of  the Fabry-Perot cavity $\Delta \omega$ using $n_g = \pi c /  L \Delta \omega$ where c is the light velocity and $L=1$mm the cavity length exhibits a very strong frequency dependence. Indeed, such dependence is expected as it arises from the ``slowing down" of the group velocity as the upper polariton departs from the light line and converges to the pure phonon excitation. Such effect is clearly apparent in Fig.~\ref{fig:spectra} b) where this dispersion is clearly shown by the uneven spacing of the cavity mode of the phonon-polariton laser that exhibits a group index varying between $5.5$ and $6.5$ in an energy range smaller than $1$meV. 

\begin{figure}[h]
  \centering
  \includegraphics[width=10cm]{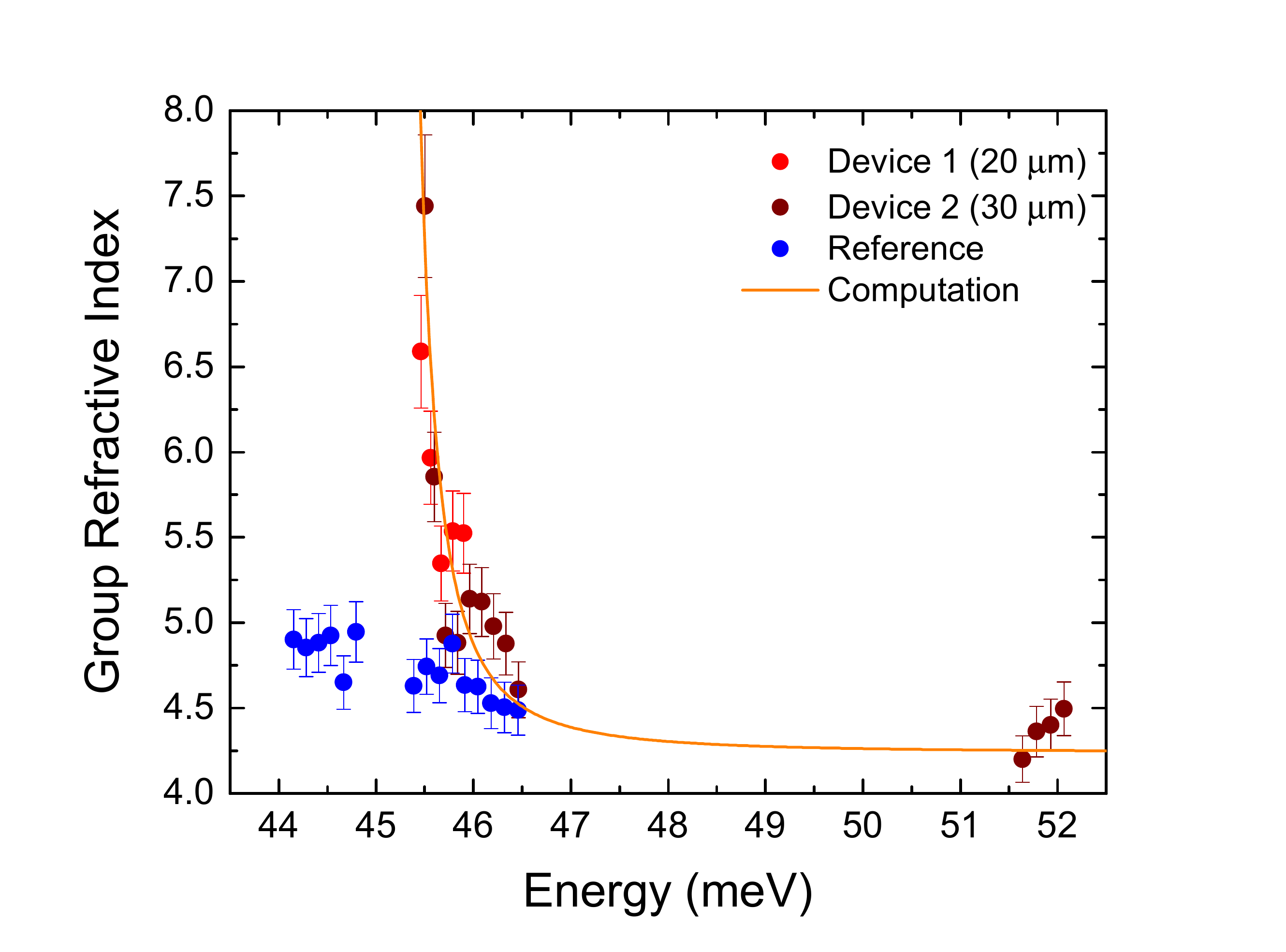}
  \caption{Comparison between the group refractive index (red and dark red disks), as derived from the optical spectrum with the one derived from the computed polariton dispersion. The group index for the reference structure (solid blue disks) is shown for comparison.}
  \label{fig:groupdispersion}
\end{figure}

The polariton dispersion can be understood by considering a model originally developed to study the ultra-strong coupling between intersubband electronic excitations and  metal-metal waveguide cavities~\cite{Todorov:2012fj}. In this model, the light-matter coupling is studied in the dipole gauge, and the quadratic $A^2$ term appears as a polarization self-interaction. We apply this model, considering the AlAs TO phonons as providing a ``mechanical" resonance at $\omega = \omega_{TO}$. In this framework, the Rabi coupling between polarization arising from the dipoles and the cavity, responsible for the anticrossing between the oscillators and the light mode, can be written at resonance as  
\begin{equation}
\Omega_R = \frac{\omega_P}{2} \sqrt{f_P},
\end{equation}
where $\omega_P$ is a plasma frequency associated with the oscillators (in our case the AlAs phonon modes) and $f_P$ their ``filling fraction". The computation of the polaritonic branches enables us to compute the group index $n_g = c/v_g = c \frac{\partial k}{ \partial \omega}$ as a function of $\omega $ and compare it to the spectral measurements of the devices, as shown in Fig.~\ref{fig:groupdispersion}. The excellent agreement between the computed dispersion and the measured one is a strong proof of the polaritonic nature of the emission. In this model, we assumed a value of $\omega_P = 14.2$meV, close to the value $\omega_P = \sqrt{\omega_{LO}^2 - \omega_{TO}^2} = 15.1$meV for AlAs bulk material. We attributed the small difference between the two values to the  slight spreading of the Aluminum atoms in the growth direction, reducing their effective local concentration. 

\begin{figure}[h]
  \centering
  \includegraphics[width=8cm]{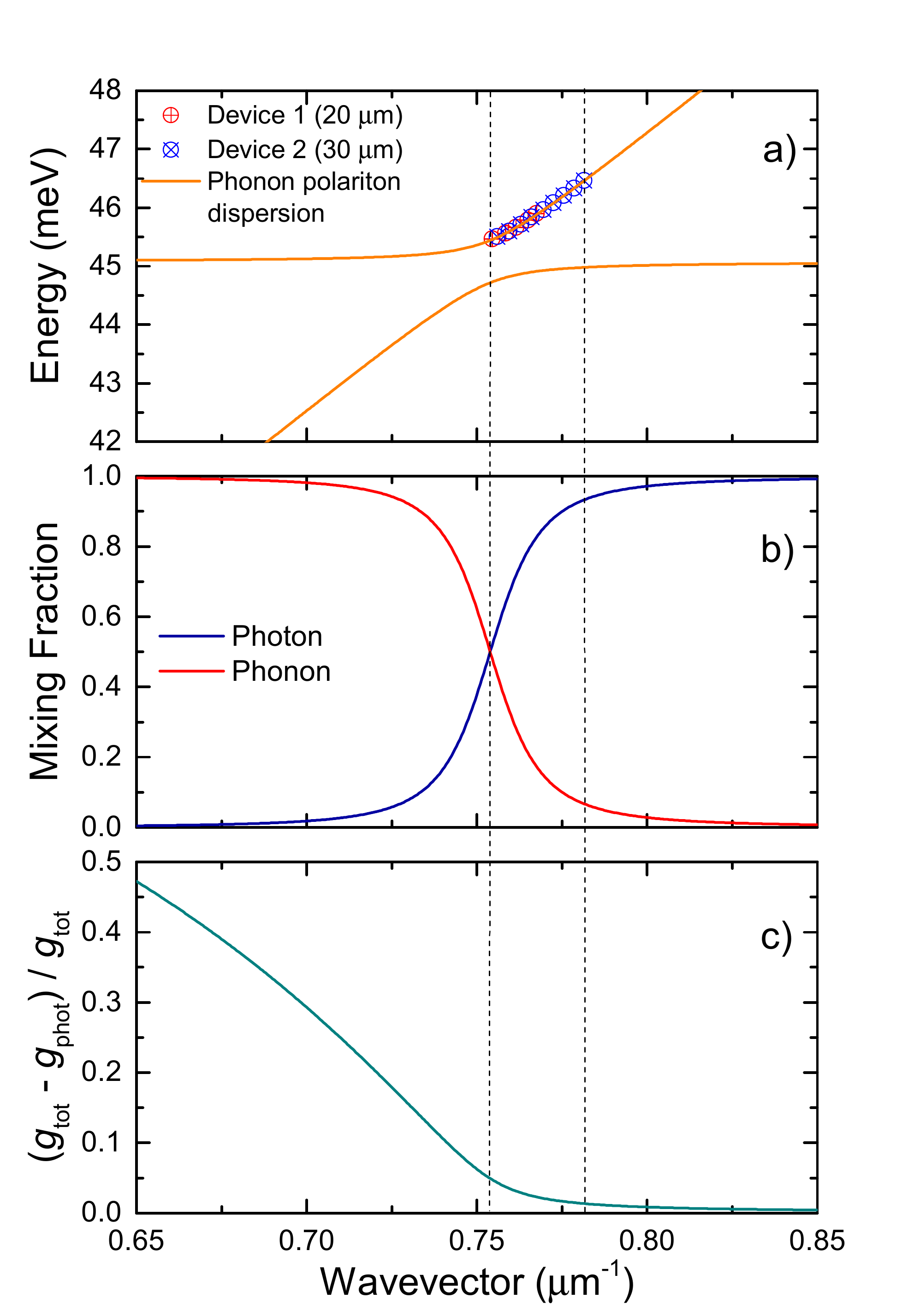}
  \caption{a) Computed dispersion of the cavity phonon-polaritons (solid lines), showing the measured spectra for two devices, as indicated. b) Computed mixing fraction of the phonon and photon part as a function of in-plane wave vector. c) Fraction of the gain arising directly from the coupling of the wavefunction to the phonons.}
  \label{fig:dispersion}
\end{figure}

Similarly, we assumed in our cavity a value of $f_P = 5.4 \times 10^{-3}$, reducing the Rabi frequency $\Omega_R$ by a factor of 14 approximately as compared to the bulk AlAs value. This value of $f_P$ would correspond to a full  monolayer of AlAs rather than a monolayer of Al$_{0.48}$In$_{0.52}$As.  As infrared absorption measurements in the restrahlen band region (shown in the supplementary material) are also consistent with this larger overall amount of AlAs material, we attributed the difference to an uncertainty in the growth process. 
In Fig.~\ref{fig:dispersion} a), the measured emission is reported onto the upper branch of the phonon-polariton dispersion curve. From their position, the phonon and photon fraction of the polariton can be retrieved from the computation of the Hopfield coefficients. As shown in Fig.~\ref{fig:dispersion} b), a phonon fraction of the polariton as high as 50\% is inferred. 

The model can then be expanded to take into account the polaritonic gain by considering the total matter polarization as the sum of the phonon ${\bf P}_{\mathrm P}$ and the intersubband ${\bf P}_{\mathrm{ISB}}$ polarizations, and treating the latter as a perturbation in a polaritonic basis using Fermi's golden rule. Using the notation of \cite{Todorov:2012fj},  the microcurrent associated with the intersubband transition is written as $\xi_{\mathrm{ISB}}(z) = \phi_{up} \partial_z \phi_{dn} - \phi_{dn} \partial_z \phi_{up}$ for the upper and lower state envelope wavefunctions  $\phi_{up,dn}(z)$. The microcurrents $\xi_P(z)$ generated by the transverse phonon polarization  in a layer at position $z_0$ have a Gaussian form since the phonon potential is harmonic
\begin{equation}
\xi_{\mathrm{P}}(z) = \frac{1}{\sigma} \sqrt{\frac{2}{\pi} } \exp \left (-\frac{(z-z_0)^2}{\sigma^2} \right ),
\end{equation}
where $\sigma = 0.16$nm is the spatial spread of the microcurrent, chosen such that 
\begin{equation}
f_{\mathrm{P}} = \left ( \frac{\int \mathcal{D}(z) \xi_{\mathrm{P}}(z) dz}{\sqrt{ \int \xi_{\mathrm{P}}^2(z) dz}} \right )^2 = 5.4 \times 10^{-3}
\end{equation}
is the value of the effective overlap used above with $\mathcal{D}(z)$ the amplitude of the optical mode in the cavity normalized such that $\int \mathcal{D}(z)^2 dz = 1$. In this approach, the gain comprises two components, one arising from the term ${\bf D} \cdot {\bf P}_{\mathrm{ISB}}$ describing the  intersubband gain, one arising from $ {\bf P}_{\mathrm{ISB}} \cdot {\bf P}_{\mathrm{P}}$, i.e. the interaction between the intersubband and the phonon polarization. The latter arises only as a result of the overlap between the phonon and intersubband microcurrents, and is equal at the anticrossing resonance to 
\begin{equation}
\Xi = \frac{\omega_{P}}{4 \omega_{\mathrm{LO}}}  \frac{\int \xi_{P}(z) \xi_{\mathrm{ISB}}(z)dz}{\sqrt{\int \xi_{\mathrm{ISB}}^2(z)dz \int \xi_{P}^2}(z)dz} \omega_{\mathrm{ISB}}
\end{equation}
while the matrix element with the intersubband is 
\begin{equation}
\Omega = \frac{1}{2}  \frac{\int \mathcal{D}(z) \xi_{\mathrm{ISB}}(z) dz}{\sqrt{ \int \xi_{\mathrm{ISB}}^2 (z)dz}}\omega_{\mathrm{ISB}}
\end{equation}
where $\omega_{\mathrm{ISB}}$  is the plasma frequency associated with the intersubband polarization and is proportional to the electron density in the upper state.  The gain will finally be written as proportional to $g \sim | x(k) \Omega + y(k) \Xi |^2 $, where $x(k)$ and $y(k)$ are the Hopfield amplitudes corresponding to the photon and phonon fractions of the polariton, respectively. The fraction of the gain not arising from the intersubband photon gain is plotted in Fig.~\ref{fig:dispersion} c), showing that this fraction rises to already 5\%, even in this structure where the phonon is confined in a single monolayer of material. 

Similarly, the polariton loss can be written in terms of the weighted content of the photon and phonon parts. Experiments with the reference laser indicate a cavity loss of $\alpha_{\mathrm{tot}} = 65 \mathrm{cm}^{-1}$\cite{Ohtani:2016vg}, corresponding to a photon lifetime of $\tau_{p} = 1.7 \mathrm{ps}$. For the lifetime of an optical phonon in GaAs, values of $\tau_{p}$ of 3.5ps at 300K and 7-8ps at 77K were found experimentally~\cite{VONDERLINDE:1980fx,KASH:1988kp} and were found consistent with a theory which assumed the decay to proceed from an anharmonic coupling to two LA phonons~\cite{Bhatt:1994eb}.

\begin{figure}[h]
  \centering
  \includegraphics[width=10cm]{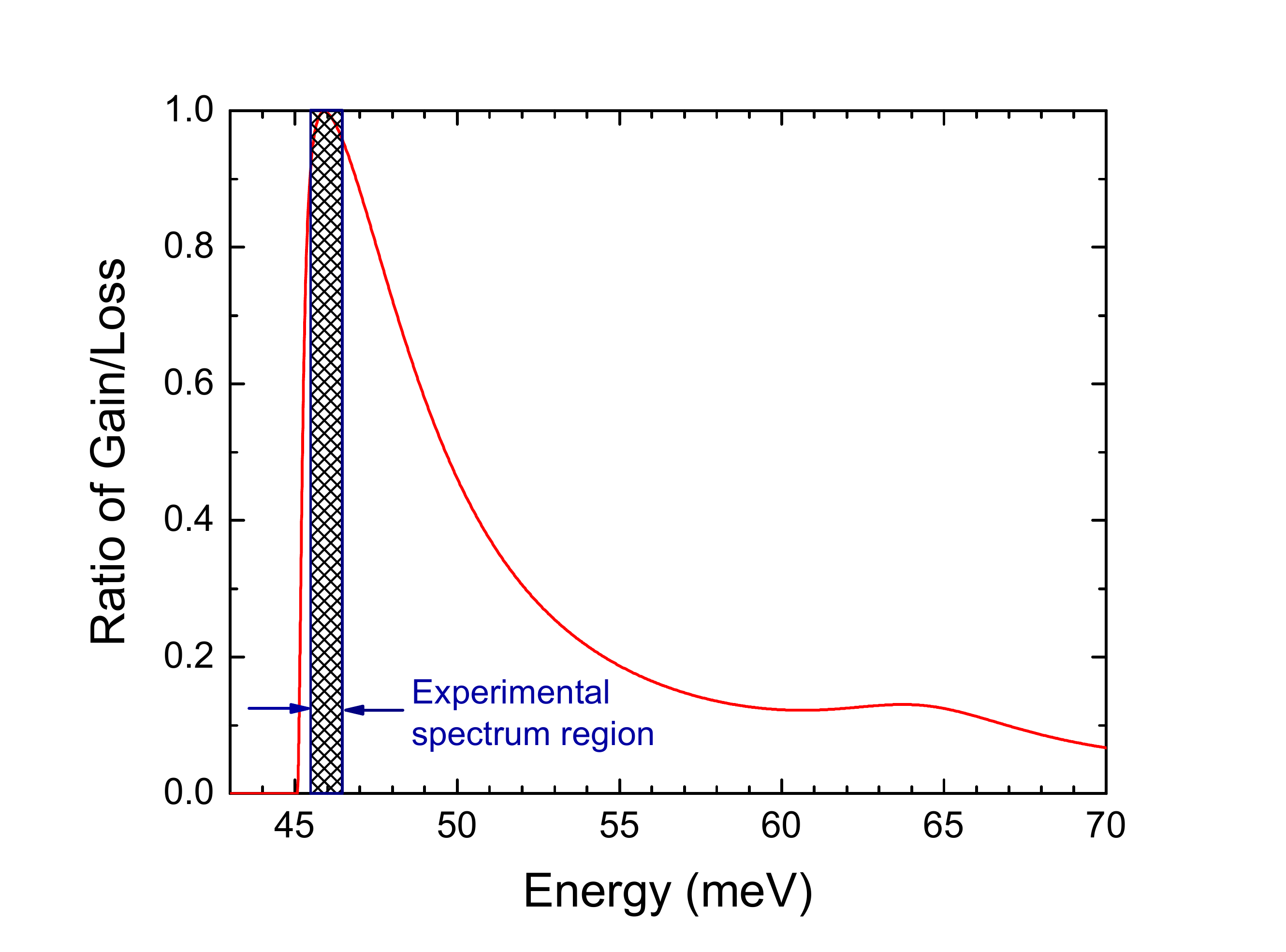}
  \caption{Computed ratio between gain and loss for the phonon-polariton laser. The observed range of laser action is overlayed on the computation. }
  \label{fig:gaincomp}
\end{figure}

 The computation of the ratio of gain to loss is shown in Fig.~\ref{fig:gaincomp}. Both gain and loss drop as $k$ decreases and the polariton becomes more phonon like, but the ratio exhibits a maximum where polariton lasing was observed. 

We attempted to observe the emission of the phonon-polaritons using a Raman spectroscopy. However, the backscattering geometry we used prevented their observation because of a very large $k$ mismatch caused by the small momentum of the polariton. Actually, the upper polariton branch is usually not seen even in the forward scattering geometry~\cite{Henry:1965tz}.

While the dispersion of the modes is a clear proof of the polaritonic nature of the emission, the spectral narrowing and intensity are witness of their coherent behavior, justifying our claim of a phonon-polariton laser. In the structure that is considered here, the gain remains dominated by the direct coupling between the electrons and the photons. This is not surprising and arises naturally as the device active region design was not optimized with the phonon contribution to the gain included. In fact, we believe that the direct coupling to the phonon that also contributes to the gain shows the radically different nature of the device and offers totally new optimization possibilities of terahertz devices especially for high temperature operation. 

Although this first demonstration relies on a III-V heterostructures, the  operation principle of this laser can be applied to all the material systems which provide simultaneous confinement of electrons and optical phonons in two dimensions. Such approach would be especially favorable for two-dimensional transition metal dichalcogenides~\cite{Wang:2012gm} structures because it allows more easily the confinement of the radiation in the direction of the heterostructure stack. As a first proof of concept, we did fabricate devices with much thinner active regions, down to 1.5$\mu m$ thick, that still operated with roughly the same threshold current (see supplementary material), demonstrating the unique capabilities of the phonon-polariton lasers to operate in subwavelength cavities.



\subsection*{Acknowledgements}
The presented work is supported in part the ERC project MUSIC as well as by the NCCR QSIT. One of us (JF) acknowledges very fruitful discussions with Angela Vasanelli, Simone de Liberato and Jacob B. Khurgin.

\newpage

\section*{Supplementary Material}

\subsection*{Device Description and band structure}
\begin{figure}[ht!]
	\centering
	\includegraphics[width=10.0cm]{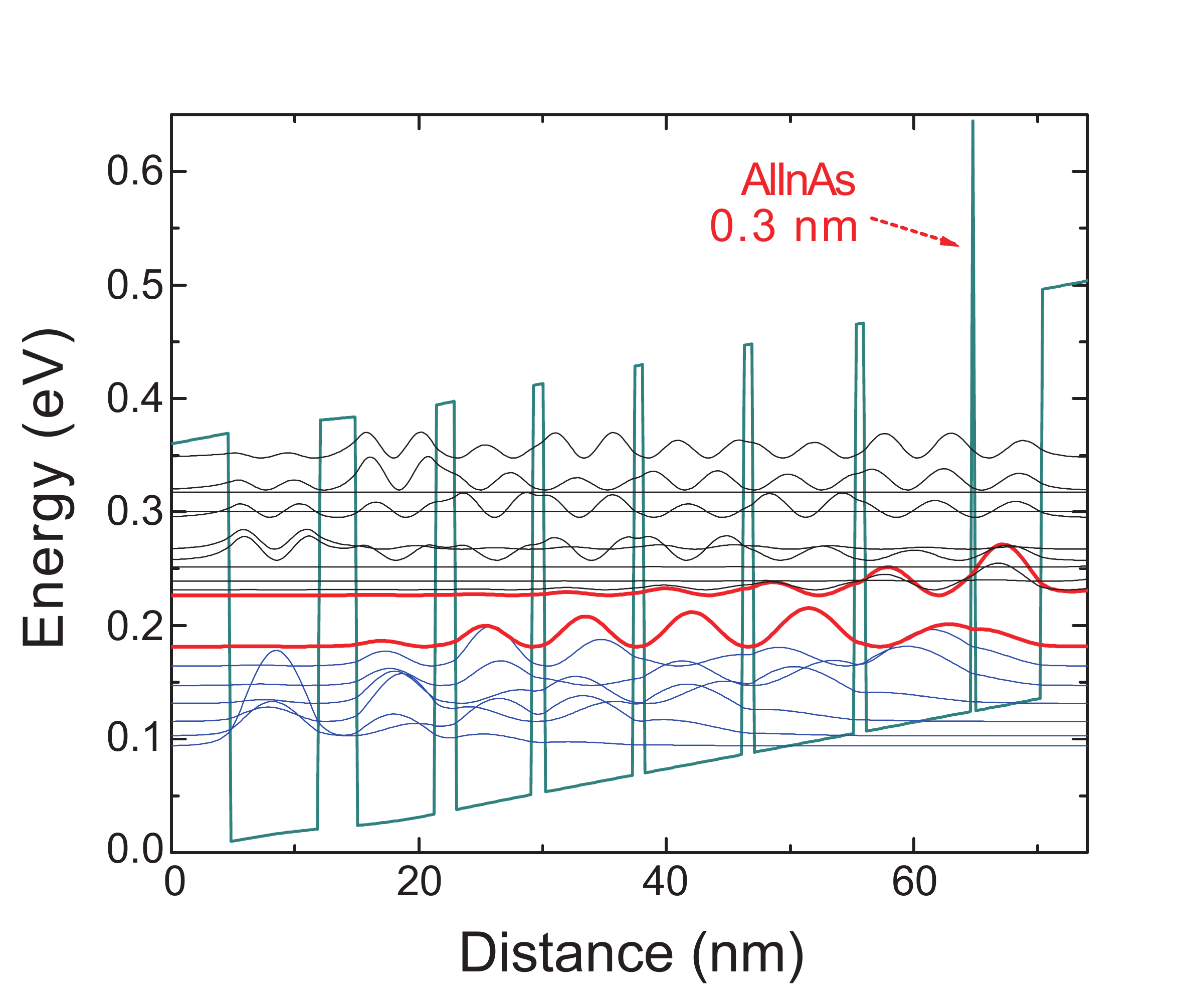}
	\caption{Conduction band diagram of one period in the active region and square modulus of the relevant wavefunctions.}
	\label{figs1}
\end{figure}

The sample was grown by molecular beam epitaxy on a Fe-doped InP substrate. The growth was started with a $50$ nm thick In$_{0.53}$Ga$_{0.47}$As buffer layer. Then a $4.2\mu$m thick active region was deposited and the growth was completed by a 30 nm thick Si-doped n-type Ga$_{0.53}$In$_{0.47}$As contact layer. In the active region, the following layer structure was repeated by $60$ times: in nanometers,  $\mathbf{4.8}/5.4/\mathit{0.3}/8.6/\mathbf{0.75}/8.2/\mathbf{0.75}/8.1/\mathbf{0.85}/7.1/\mathbf{1.13}/6.1/\mathbf{1.6}/\underline{6.4}/\mathbf{2.0}/7.2$, 
where \\GaAs$_{0.51}$Sb$_{0.49}$ barrier layers are in bold, In$_{0.53}$Ga$_{0.47}$As well layers are in roman, Al$_{0.48}$In$_{0.52}$As barrier layer is in Italic, and Si-doped In$_{0.53}$Ga$_{0.47}$As layer ($\textit{n} = 4.0 \times 10^{17}$ cm$^{-3}$) is underlined. We also grew the reference sample which is identical except for $0.55$ nm thick GaAs$_{0.51}$Sb$_{0.49}$ barrier instead of $0.3$ nm thick Al$_{0.48}$In$_{0.52}$As barrier. The corresponding conduction band diagram of the one period of the active region with moduli-squared relevant wave functions is shown in Fig.\ref{figs1}. Here an electric field of $19.5$ kV/cm is applied to align subbands. The electron effective mass in the computation is $0.043$ m$_0$ for In$_{0.53}$Ga$_{0.47}$As, $0.045$ m$_0$ for GaAs$_{0.51}$Sb$_{0.49}$ \cite{Deutsch2012} and $0.076$ m$_0$ for Al$_{0.48}$In$_{0.52}$As (m$_0$ is a free electron mass), respectively. The conduction band offset energy ($= 0.36$ eV) between In$_{0.53}$Ga$_{0.47}$As and GaAs$_{0.51}$Sb$_{0.49}$ \cite{Deutsch2012} and the one ($= 0.52$ eV) between In$_{0.5}$Ga$_{0.47}$As and Al$_{0.48}$In$_{0.52}$As are used. 

\subsection*{Device frabrication and measurements}
First a bottom metal contact layer composed of a $10$ nm thick Ti and a $500$ nm thick Au was deposited on the grown wafer. It was wafer-bonded onto the Au evaporated \textit{n}-InP (100) carrier substrate by thermos-compression bonding technique. Then the Fe-doped InP substrate used for the epitaxial growth was removed by mechanical polishing and selective wet chemical etching. After depositing Al$_2$O$_3$/SiO$_2$ layers for an electrical isolation, $250$ nm thick top contact (Ti/Au) and $800$ nm thick hard mask (Si$_3$N$_4$) layers were evaporated by electron beam evaporator. Finally the laser structures were defined by Cl$_2$-based dry etching. After device fabrication, the processed wafer was cleaved into $0.25$-$1.0$ mm long Fabry-P\'{e}rot lasers with a ridge width of $10$-$35$ $\mu$m, and then mounted on copper mounts. The laser devices were placed in a temperature controlled He flow cryostat with KRS$5$ vacuum windows.
The sub-threshold electroluminescence measurements were performed by a home-made vacuum Fourier Transform Infrared (FT-IR) spectrometer equipped with the calibrated Si bolometer. Current pulse trains composed by $100$ ns wide pulses with a repetition rate of $1$ MHz (a duty cycle of $10\%$) were modulated at $450$ Hz to match the frequency response of the Si bolometer. The light output power was also measured by the Si bolometer. For high resolution spectral measurements, Bruker Vertex $80$ vacuum FT-IR was used. The spectral resolution was $0.075$ cm$^{-1}$ ($\approx 2.2$ GHz).  

\subsection*{Far-infrared Absorption Measurements}
\begin{figure}[ht!]
	\centering
	\includegraphics[width=10.0cm]{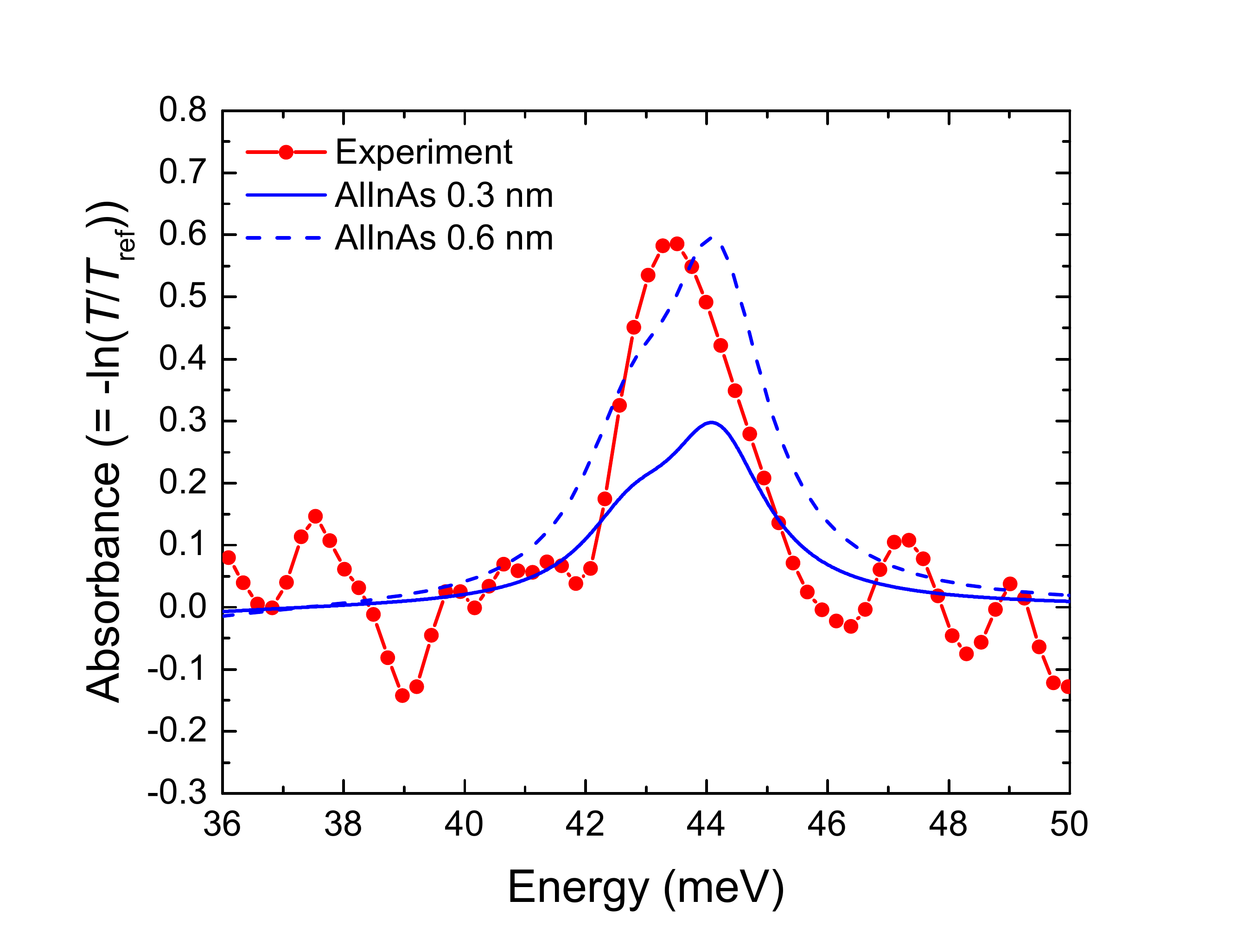}
	\caption{Room temperature absorption spectrum. Here the absorbance is defined as - ln(\textit{T}/\textit{T}$_{ref}$). The two blue curves show the computed results.} 
	\label{figs2}
\end{figure}

To probe Reststrahlen band of the AlAs optical phonon in the 0.3 nm thick AlInAs barrier, we performed far-infrared absorption measurements. In order to avoid absorption of the Fe-doped InP substrate where the Reststrahlen band is spectrally overlapped with the AlAs phonon one, the grown wafer was first glued on a Si substrate by epoxy, and mechanically polished down to $\approx 50$ $\mu$m. Then the remaining InP substrate was completely removed by a selective etching. Hydrochloric acid was used because it does not etch InGaAs at all while the etching speed of InP is fast (about 10 $\mu$m/min). Finally two edges were polished in wedge shapes with an angle of 45 degree, enabling the incident light to couple efficiently with the active region. In order to remove effects of other absorptions, we measured transmission spectrum (\textit{T}$_{ref}$) of the reference sample and then subtracted it from that (\textit{T}) of the phonon-polariton sample. Thus the absorbance is here defined as - ln(\textit{T}/\textit{T}$_{ref}$). The difference of the sample size is within 50 $\mu$m for the light propagation direction and within $2\ \mu$m for the thickness, which provides the same number (7 times) of pass across the active region for the two samples. The light was normally incident on the surface of the wedge. The red curve in Fig.\ref{figs2} shows an absorption spectrum at room temperature. An absorption peak, which is associated with the AlAs TO phonon \cite{Groenen1998}, was observed at 43.4 meV. The blue line depicts a computed absorbance. The absorption energy is in close agreement with the AlAs bulk TO phonon energy (42.8 meV) \cite{Pavesi1995}. The absorbance is about two times smaller than the computed value. As shown by the dashed blue line, the computed absorbance exhibits an agreement with the experiment if the AlInAs barrier thickness is assumed to be twice. 

\subsection*{Thinner Devices}
The larger gain of the polariton laser should translate in a capability of operating with smaller cavities. Therefore, we have investigated the dependence of the threshold on the length of the device. 
Another attempt was made by making the device thinner - rather than shorter.

For this purpose, we have processed double metal waveguide ridges of three different active region thicknesses: $4.3\ \mu$m (as-grown thickness), $3.0\ \mu$m and $1.5\ \mu$m. An $85\%$ concentration Phosphoric acid in a H$_3$PO$_4$:H$_2$O$_2$:H$_2$O ($1$:$1$:$3$) solution has been used to etch down the active region until the proper thickness was reached. Apart from the H$_3$PO$_4$ etching described above, the devices endured a standard double metal ridge processing. After this, the devices were cleaved in $0.5$ mm long dice with $26\ \mu$m wide ridges.

\begin{figure}[ht!]
	\centering
	\includegraphics[width=10.0cm]{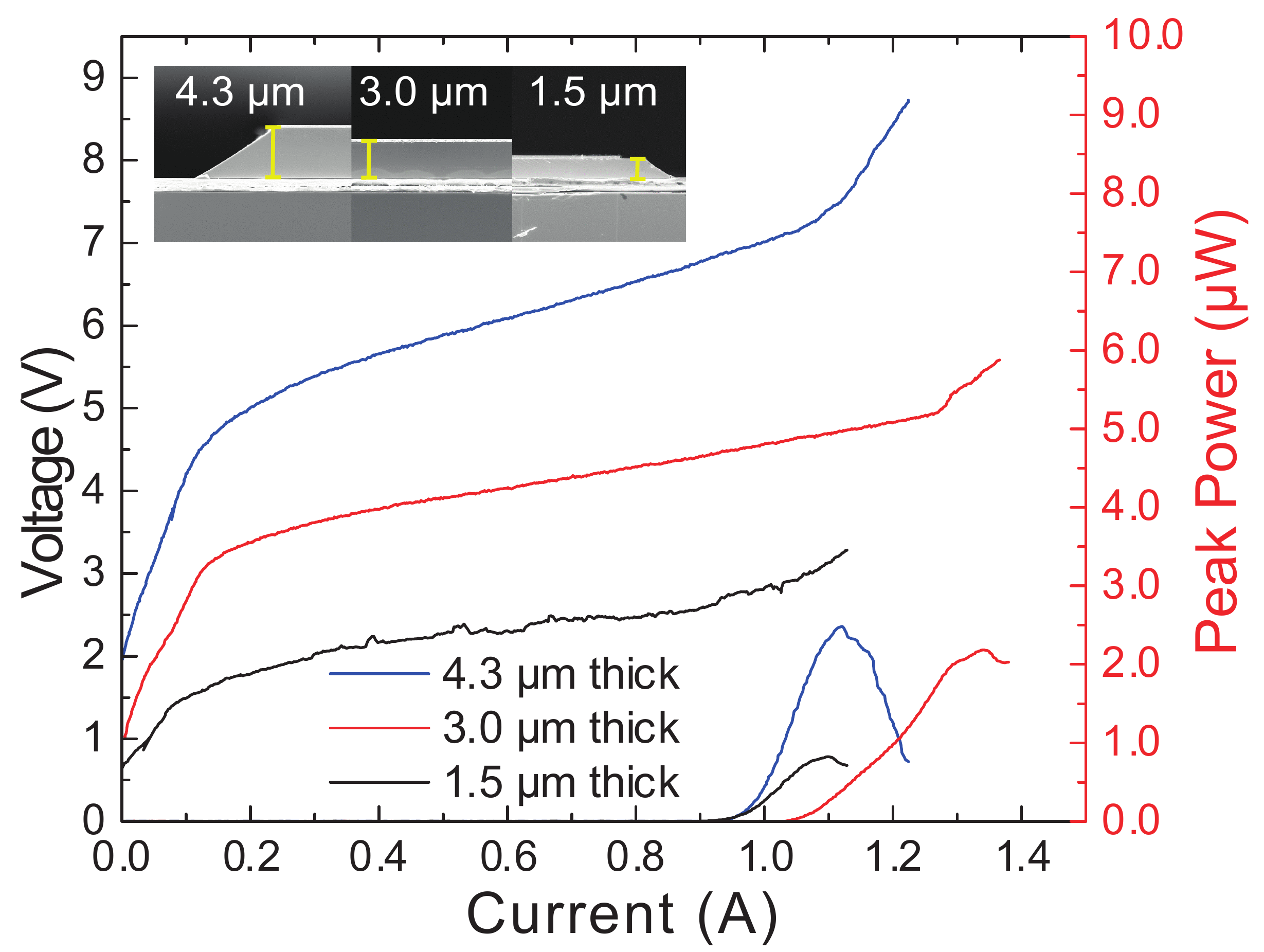}
	\caption{Light output power, current, and voltage of a phonon-polariton laser as a function of active region thickness. The inset shows a SEM cross-section of the three devices.}
	\label{fig:lithindevices}
\end{figure}

Fig. \ref{fig:lithindevices} shows the light output power, current, and voltage characteristics of these devices. Encouraging results were achieved as a working laser with active region as thin as $1.5\ \mu$m was processed, demonstrating the possibility to realize emitting structure with a subwavelength dimension. Moreover, the threshold current slightly changes ($\sim 10\%$ - $15\%$) as the thickness decreases, suggesting that it could be further decreased.

\begin{thebibliography}{10}
	
	\bibitem{VAHALA:2009ih}
	K.~Vahala, {\it et~al.\/}, {\it Nature Physics\/} {\bf 5}, 682 (2009).
	
	\bibitem{Grudinin:2010p1495}
	I.~S. Grudinin, H.~Lee, O.~Painter, K.~J. Vahala, {\it Physical Review
		Letters\/} {\bf 104}, 083901 (2010).
	
	\bibitem{Mahboob:2013dz}
	I.~Mahboob, K.~Nishiguchi, A.~Fujiwara, H.~Yamaguchi, {\it Physical Review
		Letters\/} {\bf 110}, 127202 (2013).
	
	\bibitem{Beardsley:2010de}
	R.~P. Beardsley, A.~V. Akimov, M.~Henini, A.~J. Kent, {\it Physical Review
		Letters\/} {\bf 104}, 085501 (2010).
	
	\bibitem{Tonouchi:2007p1411}
	M.~Tonouchi, {\it Nature Photonics\/} {\bf 1}, 97 (2007).
	
	\bibitem{Wolff:1970uk}
	P.~A. Wolff, {\it Physical Review Letters\/} {\bf 24}, 266 (1970).
	
	\bibitem{Fuchs:1991tw}
	F.~Fuchs, {\it et~al.\/}, {\it Physical Review Letters\/} {\bf 67}, 1310
	(1991).
	
	\bibitem{JingChen:2003jh}
	J.~Chen, J.~B. Khurgin, {\it Quantum Electronics, IEEE Journal of\/} {\bf 39},
	600 (2003).
	
	\bibitem{Spagnolo:2002p1956}
	V.~Spagnolo, {\it et~al.\/}, {\it Applied Physics Letters\/} {\bf 80}, 4303
	(2002).
	
	\bibitem{Henry:1965tz}
	C.~H. Henry, J.~J. Hopfield, {\it Physical Review Letters\/} {\bf 15}, 964
	(1965).
	
	\bibitem{Weisbuch:1992p12}
	C.~Weisbuch, M.~Nishioka, A.~Ishikawa, Y.~Arakawa, {\it Physical Review
		Letters\/} {\bf 69}, 3314 (1992).
	
	\bibitem{Imamoglu:1996un}
	A.~Imamoglu, R.~Ram, S.~Pau, Y.~Yamamoto, {\it Physical Review A\/} {\bf 53},
	4250 (1996).
	
	\bibitem{Bajoni:2008ef}
	D.~Bajoni, {\it et~al.\/}, {\it Physical Review Letters\/} {\bf 100} (2008).
	
	\bibitem{Christmann:2008p1820}
	G.~Christmann, R.~Butte, E.~Feltin, J.-F. Carlin, N.~Grandjean, {\it Applied
		Physics Letters\/} {\bf 93}, 051102 (2008).
	
	\bibitem{Schneider:2013ix}
	C.~Schneider, {\it et~al.\/}, {\it Nature\/} {\bf 497}, 348 (2013).
	
	\bibitem{Kasprzak:2006p721}
	J.~Kasprzak, {\it et~al.\/}, {\it Nature\/} {\bf 443}, 409 (2006).
	
	\bibitem{Carusotto:2013gh}
	I.~Carusotto, C.~Ciuti, {\it Reviews Of Modern Physics\/} {\bf 85}, 299 (2013).
	
	\bibitem{Faist:1994p1420}
	J.~Faist, {\it et~al.\/}, {\it Science\/} {\bf 264}, 553 (1994).
	
	\bibitem{Deutsch:2010p1906}
	C.~Deutsch, {\it et~al.\/}, {\it Applied Physics Letters\/} {\bf 97}, 261110
	(2010).
	
	\bibitem{Terazzi:2010p1513}
	R.~Terazzi, J.~Faist, {\it New Journal of Physics\/} {\bf 12}, 033045 (2010).
	
	\bibitem{Unterrainer:2002ii}
	K.~Unterrainer, {\it et~al.\/}, {\it Applied Physics Letters\/} {\bf 80}, 3060
	(2002).
	
	\bibitem{Todorov:2012fj}
	Y.~Todorov, C.~Sirtori, {\it Physical Review B\/} {\bf 85}, 045304 (2012).
	
	\bibitem{Ohtani:2016vg}
	K.~Ohtani, M.~Beck, M.~J. S{\"{u}}ess, J.~Faist, {\it arXiv:1609.08196\/}
	(2016).
	
	\bibitem{VONDERLINDE:1980fx}
	D.~Vonderlinde, J.~Kuhl, H.~Klingenberg, {\it Physical Review Letters\/} {\bf
		44}, 1505 (1980).
	
	\bibitem{KASH:1988kp}
	J.~A. Kash, J.~C. Tsang, {\it Solid-state electronics\/} {\bf 31}, 419 (1988).
	
	\bibitem{Bhatt:1994eb}
	A.~R. Bhatt, K.~W. Kim, M.~A. Stroscio, {\it Journal of Applied Physics\/} {\bf
		76}, 3905 (1994).
	
	\bibitem{Wang:2012gm}
	Q.~H. Wang, K.~Kalantar-Zadeh, A.~Kis, J.~N. Coleman, M.~S. Strano, {\it Nat.
		Nanotechnol.\/} {\bf 7}, 699 (2012).
	
\end{thebibliography}

\begin{thebibliography}{1}
	
	\bibitem{Deutsch2012}
	C.~Deutsch, {\it et~al.\/}, {\it Applied Physics Letters\/} {\bf 101}, 3
	(2012).
	
	\bibitem{Groenen1998}
	G.~Scamarcio, {\it et~al.\/}, {\it Phys. Rev. B\/} {\bf 47}, 1483 (1993).
	
	\bibitem{Pavesi1995}
	L.~Pavesi, R.~Houdr{\'e}, P.~Giannozzi, {\it Journal of Applied Physics\/} {\bf
		78}, 470 (1995).
	
\end{thebibliography}

\end{document}